\documentclass[12pt]{iopart}

\expandafter\let\csname equation*\endcsname\relax
\expandafter\let\csname endequation*\endcsname\relax 

\usepackage[english]{babel}
\usepackage{amsmath}
\usepackage{amssymb}
\usepackage{graphicx}
\usepackage{color}

\usepackage{graphicx}
\usepackage{caption}
\usepackage{subcaption}

\usepackage[utf8]{inputenc}
\usepackage{amsmath}
\usepackage{amsfonts}
\usepackage{graphicx} 
\usepackage{amsthm}
\usepackage{placeins}
\usepackage{verbatim}
\usepackage{booktabs}
\usepackage{cite}
\usepackage{color}
\usepackage{bm}
\usepackage{bbold}
\usepackage{nicefrac}
\usepackage[colorlinks=true,linkcolor=red,citecolor=blue]{hyperref}
\usepackage{comment}
\usepackage[all]{xy}
\usepackage[width=17.5cm, vscale=0.8]{geometry}
\usepackage{booktabs}
\def\Tr{\mathrm{Tr}}

\newcommand{\beq}{\begin{equation}}
\newcommand{\eeq}{\end{equation}}
\newcommand{\be}{\begin{equation}}
\newcommand{\ee}{\end{equation}}

\renewcommand{\e}{\mathrm{e}}

\let\e=\varepsilon   
\let\l=\lambda

   \let\P=\Pi

\def\Log{{\mathrm{Log}}}
\usepackage{graphicx,color}

\newcommand{\barr}{\begin{eqnarray}}
\newcommand{\earr}{\end{eqnarray}}
\let\baraccent=\= 
\renewcommand{\=}[1]{\stackrel{#1}{=}} 
\begin{document}

\title[Index of a matrix, complex logarithms, and multidimensional Fresnel integrals]{Index of a matrix, complex logarithms, and multidimensional Fresnel integrals}

\author{Pierpaolo Vivo}
\address{King's College London, Department of Mathematics, Strand, London WC2R 2LS (United Kingdom)}

\date{\today}

\begin{abstract}
We critically discuss the problem of finding the $\lambda$-index $\mathcal{N}(\lambda)\in [0,1,\ldots,N]$ of a real symmetric matrix $\bm M$, defined as the number of eigenvalues smaller than $\lambda$, using the \emph{entries} of $\bm M$ as only input. We show that a widely used formula
\begin{equation*}
\mathcal{N}(\lambda)=\lim_{\e\to 0^+}\frac{1}{2\pi \mathrm{i}}\left[\log\det(\bm M-\l+\mathrm{i}\e)-\log\det(\bm M-\l-\mathrm{i}\e)\right]
\end{equation*}
based on the branch-cut structure of the complex logarithm should be handled with care, as it generically fails to produce the correct result if the same branch is chosen for the two logarithms. 

We improve the formula using multidimensional Fresnel integrals, showing that even the new version provides at most a self-consistency equation for $\mathcal{N}(\lambda)$, whose solution is not guaranteed to be unique. Our results are corroborated by explicit examples and numerical evaluations.

\end{abstract}

\maketitle


\section{The question}
Consider a $N\times N$ real symmetric matrix $\bm M = (M_{ij})$, whose eigenvalues (all real) are $\lambda_1\leq\lambda_2\leq\ldots\leq \lambda_N$. Is there a way to count how many eigenvalues of $\bm M$ fall below a threshold $\lambda$, using as only input\footnote{In particular, explicit diagonalisation and Sylvester's law of inertia are out of the game.} the \emph{entries} of $\bm M$? This question might appear rather bizarre at first sight. In practice, though, this is often the scenario one faces when dealing with non-invariant \emph{random} matrices, where the only available information concerns the distribution of matrix \emph{entries}. This setting is relevant to Wigner matrices, especially in connection with sparse undirected graphs. Having a formula that connects directly the entries of $\bm M$ to the number of its eigenvalues falling below a certain threshold - an interesting and well-studied observable in the context of random matrices \cite{scard1,scard2} - would certainly come very handy in these cases.

For the time being, let us focus on a single, deterministic matrix $\bm M$, and define the \emph{$\lambda$-index} of $\bm M$, denoted by $\mathcal{N}(\lambda)$ as
\begin{equation}
\mathcal{N}(\lambda)=\# \text{ of eigenvalues of }\bm M\text{ less than } \lambda\ .
\end{equation}

A formula that is claimed to do the job exists (see \cite{cavagna,metz,metzreplica,castillo,fyodorov} for equivalent or related versions of the formula given here) 
\begin{equation}
\mathcal{N}(\lambda)=\lim_{\e\to 0^+}\frac{1}{2\pi \mathrm{i}}\left[\log\det(\bm M-\l+\mathrm{i}\e)-\log\det(\bm M-\l-\mathrm{i}\e)\right]\ .\label{mainformula}
\end{equation}

At first sight, this formula indeed seems like an excellent candidate. First, it only requires as input the \emph{entries} of $\bm M$ (in the form of a determinant). Second, the \emph{complex} function $f(z)=\log\det(\bm M - z)$, having cuts on the real axis, allegedly counts how many eigenvalues fall below $\lambda$ by recording how many jumps of width $2\pi \mathrm{i}$ one meets while approaching the real axis from above and from below. But is this really the case?

The first thing to notice is that the notation $\log$ ($\ln$ is also often used) in \eqref{mainformula} is ambiguous: the complex logarithm is a multi-valued function (see \ref{appA} for details), hence it is important to specify which branch is chosen. \emph{If} the same branch is chosen for the two logarithms appearing in \eqref{mainformula} (for example, the Principal branch, the default choice in \textsf{Mathematica}), then a simple example will be sufficient to show that $\mathcal{N}(\l)$ can only take a handful of (half)-integer values, making \eqref{mainformula} rather unfit for purpose.

\subsection{Problem n. 1}

Take
\be
\bm M=
\left(
\begin{array}{cccc}
 -1 & -2 & 1 & 3 \\
 -2 & -4 & 2 & 2 \\
 1 & 2 & -4 & -5 \\
 3 & 2 & -5 & -3 \\
\end{array}
\right)\ ,
\ee
whose eigenvalues are $\bm\lambda=\{\lambda_1,\lambda_2,\lambda_2,\lambda_4\}=\{-11.0323...,-2.79789...,-0.631656...,2.46186...\}$. Therefore, $\mathcal{N}(0)=3$ (just an example for $\lambda=0$). Now, let us pretend we do not have any way to access the list of eigenvalues directly (hence we cannot just count them!). We just wish to infer $\mathcal{N}(0)$ by performing manipulations on the entries of $\bm M$ alone.

Trying to use \eqref{mainformula}, we have to compute
\begin{align}
\det(\bm M+\mathrm{i}\epsilon)&=\epsilon^4+12 \mathrm{i} \epsilon^3-4 \epsilon^2+78 \mathrm{i} \epsilon-48\\
\det(\bm M-\mathrm{i}\epsilon)&=\epsilon^4-12 \mathrm{i} \epsilon^3-4 \epsilon^2-78 \mathrm{i} \epsilon-48\ ,
\end{align}
whose principal logarithms (hereafter denoted by $\Log$), evaluated e.g. for $\epsilon=10^{-8}$ in \textsf{Mathematica} yield
\begin{align}
\Log\det(\bm M+\mathrm{i}\epsilon)&\simeq 3.871201010907891 + 3.141592637339793\mathrm{i}\\
\Log\det(\bm M-\mathrm{i}\epsilon)&\simeq 3.871201010907891 - 3.141592637339793\mathrm{i}\ .
\end{align}
The difference between the two is quite clearly $2\pi\mathrm{i}$, therefore $\mathcal{N}(0)=1$ using \eqref{mainformula} (and not $\mathcal{N}(0)=3$ as expected!). This is understandable, as the difference between imaginary parts of two principal logarithms is bounded between $-2 \pi\mathrm{i}$ and $2\pi\mathrm{i}$, being therefore rather unfit for any counting purpose.

We can try now to gain some further intuition by exploiting the very bit of information that we pledged \emph{not} to use, namely the exact values $\bm\lambda=\{\lambda_1,\lambda_2,\lambda_2,\lambda_4\}$ of the eigenvalues of $\bm M$. Let us cheat for a moment then, and compute numerically the following quantities to machine precision (fix e.g. $\epsilon=10^{-8}$)
\begin{align}
\Log(\lambda_1+\mathrm{i}\epsilon) &\simeq 2.400829079775955 + 3.1415926526833657\mathrm{i}\label{a3}\\
\Log(\lambda_1-\mathrm{i}\epsilon) &\simeq 2.400829079775955 - 3.1415926526833657\mathrm{i}\label{a4}\\
\Log(\lambda_4+\mathrm{i}\epsilon) &\simeq 0.9009181991462066 + 10^{-9} \mathrm{i}\\
\Log(\lambda_4-\mathrm{i}\epsilon) &\simeq 0.9009181991462066 - 10^{-9} \mathrm{i}\ .
\end{align}
For each of the \emph{negative} eigenvalues (like $\lambda_1$), subtracting \eqref{a4} from \eqref{a3} we get a contribution $2\pi\mathrm{i}$, while for each of the \emph{positive} eigenvalues (like $\lambda_4$) there is no imaginary part and the subtraction gives $0$. Therefore we get as many $2\pi\mathrm{i}$ contributions as there are negative eigenvalues (or more generally, eigenvalues $<\lambda $). But isn't this precisely achieving the counting of eigenvalues that we were after? Indeed, the following formula does the job perfectly\footnote{Not completely true, but bear with us for a moment.}
\be
\mathcal{N}(\lambda) = \lim_{\epsilon\to 0^+}\frac{1}{2\pi \mathrm{i}}\left[\sum_{i=1}^N \Log(\lambda_i-\lambda+\mathrm{i}\epsilon)-\sum_{i=1}^N \Log(\lambda_i-\lambda-\mathrm{i}\epsilon)\right]\ ,\label{formularight}
\ee
and in our case would correctly return $\mathcal{N}(0)=3$. The point is that \eqref{formularight} is \emph{not} equivalent to \eqref{mainformula}, as one might be naively tempted to assume (using the standard \emph{mantra} $\Tr\log = \log\det$).

Let us now pause for a moment and summarise what we have done. We have shown that \emph{knowing the individual eigenvalues} we can determine $\mathcal{N}(\lambda)$\footnote{Obviously, we are purposely using an overly complicated method, instead of just counting them!} exploiting the branch cut structure of the complex logarithm, and land on the correct answer (Eq. \eqref{formularight}). If instead we do \emph{not} know (or cannot use) the eigenvalues (as postulated at the beginning), and we try to cast Eq. \eqref{formularight} in the form of Eq. \eqref{mainformula} (involving only the \emph{entries} of $\bm M$), we get the wrong answer. Why?

The reason can be ultimately traced back to the failure of the standard relation among \emph{real} variables $\ln(x_1 x_2)=\ln(x_1)+\ln(x_2)$ in a \emph{complex} setting. Indeed, in general
\be
\Log(z_1 z_2) \neq \Log(z_1) +\Log(z_2)\ ,
\ee
as the imaginary part of the l.h.s. is bounded between $-\pi$ and $\pi$, while the imaginary part of the r.h.s. - being the \emph{sum} of two - is not guaranteed to lie again between $-\pi$ and $\pi$. That is why the sum of principal logarithms in \eqref{formularight} does not translate into a single principal logarithm of a product (the determinant in Eq. \eqref{mainformula}).

Would fiddling with the branches of the complex logarithm in Eq. \eqref{mainformula} cure the problem?

Choosing the \emph{same} (non-principal) branch for the two logarithms appearing in Eq. \eqref{mainformula} will unfortunately have no effect: the addition of the \emph{same} amount $2\pi\mathrm{i}n$ ($n\in\mathbb{N}$) to each of the two principal values (which is the operation needed to define a new non-principal logarithm, see \ref{appA})
gets wiped out in the subtraction. One would be forced to choose two \emph{different} branches of the complex logarithms depending on the specific form of the matrix $\bm M$ and the value of the threshold $\lambda$, with the obvious drawback that the resulting formula will be heavily case-dependent and ultimately of little use.

\subsection{Problem n. 2}

There is yet another problem with this formalism. Indeed, Eq. \eqref{formularight} is not completely flawless either.

The problem specifically arises when the matrix $\bm M$ has $N_0$ eigenvalues \emph{exactly} equal to $\l$. Then, applying \eqref{formularight} we would get
\begin{align}
&\mathcal{N}(\l) \stackrel{?}{=}
 \lim_{\epsilon\to 0^+}\frac{1}{2\pi \mathrm{i}}\left[\sum_{i=1}^N \Log(\lambda_i-\l+\mathrm{i}\epsilon)-\sum_{i=1}^N \Log(\lambda_i-\l-\mathrm{i}\epsilon)\right]\\
 &= \lim_{\epsilon\to 0^+}\frac{1}{2\pi \mathrm{i}}\left[N_0 \left(\Log(\mathrm{i}\epsilon)-\Log(-\mathrm{i}\epsilon)\right)+\left(\sum_{i:\lambda_i\neq \l} \Log(\lambda_i-\l+\mathrm{i}\epsilon)-\sum_{i:\lambda_i\neq \lambda}^N \Log(\lambda_i-\l-\mathrm{i}\epsilon)\right)\right]\ ,\label{formulawrong}
\end{align}
where we have isolated the contribution from the eigenvalues that are exactly equal to $\l$. However, due to the following identity (proved in \ref{appC})
\be
\lim_{\epsilon\to 0^+}\frac{1}{2\pi\mathrm{i}}\left[\Log(\mathrm{i}\epsilon)-\Log(-\mathrm{i}\epsilon)\right]=\frac{1}{2}\ ,\label{identity}
\ee
the formula \eqref{formularight} actually \emph{overestimates} the correct number of eigenvalues strictly smaller than $\l$ by an amount $N_0/2$.

An improved index formula can be nevertheless defined by discounting half of the number of eigenvalues exactly equal to $\l$. This is achieved by (see again \ref{appC})
\be
\begin{split}
\mathcal{N}(\l)&=\lim_{\epsilon\to 0^+}\frac{1}{2\pi\mathrm{i}}\left[\sum_{i=1}^N\left(\Log (\l_i-\l+\mathrm{i}\epsilon)-\Log(\l_i-\l-\mathrm{i}\epsilon)\right)\right.\\
&\left.-2\sum_{i=1}^N\left(\Log(\l_i-\l+\mathrm{i}\epsilon-\epsilon)-\Log(\l_i-\l+\mathrm{i}\epsilon+\epsilon)\right)\right]\ .\label{correctfinal}
\end{split}
\ee
Let us check it on the following example.
We consider
\beq
\bm M=
\left(
\begin{array}{cccc}
 -\frac{3}{2} & \frac{1}{2} & 0 & 0 \\
 \frac{1}{2} & -\frac{3}{2} & 0 & 0 \\
 0 & 0 & 2 & 0 \\
 0 & 0 & 0 & 0
\end{array}
\right)\:.
\eeq
This matrix has the following eigenvalues $\bm \l=\{\l_1,\l_2,\l_3,\l_4\}=\{-2,-1,0,2\}$.
If we apply again \eqref{formularight} to $\lambda=0$, we would get $\mathcal N(0)=5/2$, which is clearly incorrect.
Indeed we get for $\epsilon=10^{-2}$
\begin{align*}
\Log(\lambda_1+\mathrm{i}\epsilon) &\simeq 0.69 + 3.14 \mathrm{i}\qquad
&\Log(\l_1-\mathrm i \epsilon) &\simeq 0.69 - 3.14 \mathrm i\\
\Log(\lambda_2+\mathrm{i}\epsilon) &\simeq 3.14 \mathrm{i}\qquad
&\Log(\l_2-\mathrm i \epsilon) &\simeq -3.14 \mathrm i\\
\Log(\lambda_3+\mathrm{i}\epsilon) &\simeq -18.42 + 1.57 \mathrm{i}\qquad
&\Log(\lambda_3-\mathrm{i}\epsilon) &\simeq -18.42 - 1.57  \mathrm{i}\\
\Log(\lambda_4+\mathrm{i}\epsilon) &\simeq 0.69\qquad
&\Log(\lambda_4-\mathrm{i}\epsilon) &\simeq 0.69\ .
\end{align*}
Clearly the zero eigenvalue $\lambda_3$ induces a jump of width $\pi\mathrm{i}$, which interferes with the counting of eigenvalues \emph{strictly} smaller than $\lambda=0$.

Let us now consider the correction term that appears in \eqref{correctfinal}.
We get
\begin{align*}
\Log(\lambda_1+\mathrm{i}\epsilon-\epsilon) &\simeq 0.69 +3.14 \mathrm{i}\qquad
&\Log(\l_1+\mathrm i \epsilon+\epsilon)&\simeq 0.69 +3.14 \mathrm i\\
\Log(\lambda_2+\mathrm{i}\epsilon-\epsilon) &\simeq +3.14  \mathrm{i}\qquad
&\Log(\l_2+\mathrm i \epsilon+\epsilon)&\simeq +3.14  \mathrm i\\
\Log(\lambda_3+\mathrm{i}\epsilon-\epsilon) &\simeq -18.07 + 2.3 \mathrm{i}\qquad
&\Log(\lambda_3+\mathrm{i}\epsilon+\epsilon) &\simeq -18.07 + 0.78 \mathrm{i}\\
\Log(\lambda_4+\mathrm{i}\epsilon-\epsilon) &\simeq 0.69\qquad
&\Log(\lambda_4+\mathrm{i}\epsilon+\epsilon) &\simeq 0.69\ .
\end{align*}
Therefore
\begin{equation}
\Log(\lambda_3+\mathrm{i}\epsilon-\epsilon) -\Log(\lambda_3+\mathrm{i}\epsilon+\epsilon) =\frac{\pi}{2}\mathrm{i}\ ,
\end{equation}
and from \eqref{correctfinal}
\begin{equation}
\mathcal{N}(\lambda=0)=\frac{1}{2\pi\mathrm{i}}\left[2\times 2\pi\mathrm{i}+\mathrm{i}\pi-2\times\frac{\pi}{2}\mathrm{i}\right]=2
\end{equation}
as expected.
\vspace{50pt}\\
To summarise, we have produced two formulas for the $\lambda$-index (Eq.  \eqref{formularight} and \eqref{correctfinal}), which return the correct result\footnote{Modulo the incorrect handling by Eq. \eqref{formularight} of matrices with eigenvalues exactly equal to $\lambda$.}. From now on, we will focus on the ``shorter" version (Eq. \eqref{formularight}) for illustrative purposes: all our considerations can be easily extended to the more accurate version (Eq. \eqref{correctfinal}) if needed.

We have also showed that Eq. \eqref{formularight} is \emph{not} equivalent to the starting point in Eq. \eqref{mainformula}. This is really disappointing, though, because Eq. \eqref{formularight} relies on the complete knowledge of all the individual eigenvalues (which of course makes the whole exercise rather pointless), while Eq. \eqref{mainformula} does not (but fails to produce the correct result!).

Is there a way to convert the ``correct" formula \eqref{formularight} into an expression involving only the entries of the matrix $\bm M$? A possible strategy - based on a careful evaluation of multidimensional Fresnel integrals - and its limitations is outlined in the next section.

\section{Multidimensional Fresnel integrals}

For reasons that will become readily apparent, we now carefully evaluate the multidimensional Fresnel integral
\begin{equation}
\mathcal{I}_N[\bm M;\lambda;\epsilon]=\int_{(-\infty,\infty)^N}\mathrm{d}\bm x\exp\left[-\frac{\mathrm{i}}{2}\bm x^T \left( (\lambda-\mathrm{i}\epsilon)\bm{1} -\bm{M}\right)\bm x\right]\ ,\label{multidimensionalFresnel}
\end{equation}
where $\epsilon>0$, $\bm x = (x_1,\ldots,x_N)^T$ is a real column vector, $\bm M$ is a real symmetric matrix with real eigenvalues $\{\lambda_i\}$, $\lambda$ a real number and $\bm{1}$ is the $N\times N$ identity matrix. Note that the integral is perfectly convergent, as the coefficient of terms $\sim x^2$ in the exponent is \emph{negative} ($-\epsilon/2$).

As a warm-up, we consider the single Fresnel integral
\begin{equation}
\mathcal{I}_1[m;\lambda;\epsilon]=\int_{-\infty}^\infty\mathrm{d}x\exp\left[-\left(\frac{1}{2}\epsilon+\frac{\mathrm{i}}{2}(\lambda-m)\right)x^2\right]\ ,
\end{equation}
which is equal to
\begin{equation}
 \mathcal{I}_1[m;\lambda;\epsilon] =\sqrt{2\pi}\exp\left[-\frac{1}{2}\mathrm{Log}(\epsilon+\mathrm{i}(\lambda-m))\right]=\sqrt{2 \pi } \exp \left[-\frac{1}{2} \left(\mathrm{Log} (m-\lambda+\mathrm{i} \epsilon )-\frac{\pi  \mathrm{i}}{2}\right)\right]\ .\label{Isingle}
\end{equation}
Note that we have intentionally avoided writing the result in terms of square roots of complex numbers. The way the result is written in \eqref{Isingle} presents no ambiguities whatsoever, and can be easily checked numerically with arbitrary precision.

Now, turning back to \eqref{multidimensionalFresnel}, let $\bm M = \bm O \bm D \bm O^T$ be the spectral decomposition of $\bm M$, with $\bm O$ the orthogonal matrix of eigenvectors. Making a change of variable $\bm y = \bm O^T \bm x$ with unit Jacobian, we readily realize than
\begin{equation}
\mathcal{I}_N[\bm M;\lambda;\epsilon]=\int_{(-\infty,\infty)^N}\mathrm{d}\bm x\exp\left[-\frac{\mathrm{i}}{2}\bm x^T \left(\lambda\bm{1}-\bm M-\mathrm{i}\epsilon\right)\bm x\right]=\prod_{k=1}^N \mathcal{I}_1[\lambda_k;\lambda;\epsilon]\ ,\label{multidimensionalFresnel2}
\end{equation}
where $\{\lambda_k\}$ are the eigenvalues of $\bm M$.

Therefore
\begin{equation}
\mathcal{I}_N[\bm M;\lambda;\epsilon]=(2\pi)^{N/2}\exp\left[\underbrace{-\frac{1}{2}\sum_{k=1}^N\mathrm{Log}(\lambda_k-\lambda+\mathrm{i}\epsilon)+\mathrm{i}\frac{N\pi}{4}}_{\blacksquare}\right]\ .\label{identityFresnel}
\end{equation}

The first question to address is whether this expression may be cast in a more familiar (at least to a physicist's eye) form involving the inverse square root of a determinant. This procedure requires again a very meticulous care in the manipulations, and is reported in \ref{appD}.

Multiplying by $(2\pi)^{-N/2}$ and taking the principal logarithm on both sides of \eqref{identityFresnel}, we get
\begin{equation}
\mathrm{Log}[(2\pi)^{-N/2}\mathcal{I}_N[\bm M;\lambda;\epsilon]]=-\frac{1}{2}\underbrace{\sum_{k=1}^N\mathrm{Log}(\lambda_k-\lambda+\mathrm{i}\epsilon)}_{ \blacklozenge}+\mathrm{i}\frac{N\pi}{4}+2\pi\mathrm{i}\Big\lfloor \frac{1}{2}-\frac{1}{2\pi}\mathrm{Im}(\blacksquare)\Big\rfloor\ ,\label{right1}
\end{equation}
where $\lfloor\cdot\rfloor$ denotes the largest integer less than or equal to $(\cdot)$. The last term on the r.h.s. is due to another startling property of complex logarithms, namely $\mathrm{Log}(\exp(z))$ may not be just equal to $z$ ! Eq. \eqref{right1} is particularly interesting, as it connects an expression that depends on the \emph{entries} of $\bm M$ (on the l.h.s.) to the term $\blacklozenge$ (depending on the \emph{eigenvalues} of $\bm M$), which appears in \eqref{formularight}. May \eqref{right1} be then the long-sought key for our problem? It would certainly be, were it not for the last term in \eqref{right1}, which (as we now proceed to show) \emph{still} depends explicitly on the exact pattern of eigenvalues.

Indeed, by definition of principal logarithm
\begin{equation}
\mathrm{Im}(\blacksquare)=-\frac{1}{2}\sum_{k=1}^N \mathrm{atan2}(\epsilon,\lambda_k-\lambda)+\frac{N\pi}{4}\ ,\label{imaginarysquare}
\end{equation}
where $\mathrm{atan2}(y,x)$ is defined in \eqref{atan}. Since $\epsilon>0$, we have 
\begin{align}
 \mathrm{atan2}(\epsilon,\lambda_k-\lambda) &=\arctan\left(\frac{\epsilon}{\lambda_k-\lambda}\right)+\pi &\text{ if }
 \lambda_k-\lambda<0\label{crop1}\\
  \mathrm{atan2}(\epsilon,\lambda_k-\lambda) &=\arctan\left(\frac{\epsilon}{\lambda_k-\lambda}\right) &\text{ if }
 \lambda_k-\lambda>0\ .
\end{align}
In view of \eqref{crop1}, there will be as many $+\pi$ terms arising in \eqref{imaginarysquare} - once written in terms of arctan - as there are eigenvalues smaller than $\lambda$ ($=\mathcal{N}(\lambda)$).

Similarly, we may consider the multidimensional Fresnel integral
\begin{equation}
\mathcal{J}_N[\bm M;\lambda;\epsilon]=\int_{(-\infty,\infty)^N}\mathrm{d}\bm x\exp\left[\frac{\mathrm{i}}{2}\bm x^T \left( \lambda\bm{1} -\bm{M}+\mathrm{i}\epsilon\right)\bm x\right]\ ,\label{multidimensionalFresnelJ}
\end{equation}
where $\epsilon>0$, $\bm x = (x_1,\ldots,x_N)^T$ is a real column vector, $\bm M$ is a real symmetric matrix with real eigenvalues $\{\lambda_i\}$, $\lambda$ a real number and $\bm{1}$ is the $N\times N$ identity matrix. Note the change in sign of $\mathrm{i}/2$ in the argument of the exponential, necessary to ensure that the integral is convergent.

As a warm-up, we again consider the single-integral version of it
\begin{equation}
\mathcal{J}_1[m;\lambda;\epsilon]=\int_{-\infty}^\infty\mathrm{d}x\exp\left[-\left(\frac{1}{2}\epsilon-\frac{\mathrm{i}}{2}(\lambda-m)\right)x^2\right]\ ,
\end{equation}
which is equal to
\begin{equation}
\mathcal{J}_1[m;\lambda;\epsilon] =\sqrt{2\pi}\exp\left[-\frac{1}{2}\mathrm{Log}(\epsilon-\mathrm{i}(\lambda-m))\right]=\sqrt{2 \pi } \exp \left[-\frac{1}{2} \left(\mathrm{Log} (m-\lambda-\mathrm{i} \epsilon )+\frac{\pi  \mathrm{i}}{2}\right)\right]\ .
\end{equation}

Now, turning back to \eqref{multidimensionalFresnelJ}, let $\bm M = \bm O \bm D \bm O^T$ the spectral decomposition of $\bm M$, with $\bm O$ the orthogonal matrix of eigenvectors. Making a change of variable $\bm y = \bm O^T \bm x$ with unit Jacobian, we readily realize than
\begin{equation}
\mathcal{J}_N[\bm M;\lambda;\epsilon]=\int_{(-\infty,\infty)^N}\mathrm{d}\bm x\exp\left[\frac{\mathrm{i}}{2}\bm x^T \left(\lambda\bm{1}-\bm M+\mathrm{i}\epsilon\right)\bm x\right]=\prod_{k=1}^N \mathcal{J}_1[\lambda_k,\lambda,\epsilon]\ ,\label{multidimensionalFresnel2}
\end{equation}
where $\{\lambda_k\}$ are the eigenvalues of $\bm M$.

Therefore
\begin{equation}
\mathcal{J}_N[\bm M;\lambda;\epsilon]=(2\pi)^{N/2}\exp\left[\underbrace{-\frac{1}{2}\sum_{k=1}^N\mathrm{Log}(\lambda_k-\lambda-\mathrm{i}\epsilon)-\mathrm{i}\frac{N\pi}{4}}_{\blacktriangle}\right]\ .
\end{equation}
Note that the sum of logarithms appearing on the r.h.s. is precisely the second term on the r.h.s. of \eqref{formularight}.
Multiplying by $(2\pi)^{-N/2}$ and taking the principal logarithm on both sides, we get
\begin{equation}
\mathrm{Log}[(2\pi)^{-N/2}\mathcal{J}_N[\bm M;\lambda;\epsilon]]=-\frac{1}{2}\sum_{k=1}^N\mathrm{Log}(\lambda_k-\lambda-\mathrm{i}\epsilon)-\mathrm{i}\frac{N\pi}{4}+2\pi\mathrm{i}\Big\lfloor \frac{1}{2}-\frac{1}{2\pi}\mathrm{Im}(\blacktriangle)\Big\rfloor\ .\label{right2}
\end{equation}

We now proceed to evaluate $\mathrm{Im}(\blacktriangle)$. We have again by definition of principal logarithm
\begin{equation}
\mathrm{Im}(\blacktriangle)=-\frac{1}{2}\sum_{k=1}^N \mathrm{atan2}(-\epsilon,\lambda_k-\lambda)-\frac{N\pi}{4}\ ,
\end{equation}
where $\mathrm{atan2}(y,x)$ is defined in \eqref{atan}. Since $-\epsilon<0$, we have 
\begin{align}
 \mathrm{atan2}(-\epsilon,\lambda_k-\lambda) &=\arctan\left(\frac{-\epsilon}{\lambda_k-\lambda}\right)-\pi &\text{ if }
 \lambda_k-\lambda<0\label{crop2}\\
  \mathrm{atan2}(-\epsilon,\lambda_k-\lambda) &=\arctan\left(\frac{-\epsilon}{\lambda_k-\lambda}\right) &\text{ if }
 \lambda_k-\lambda>0\ .
\end{align}

Therefore, from \eqref{formularight} and using \eqref{right1} and \eqref{right2} we can write an equation for the index of the matrix $\bm M$\footnote{This formula - being derived from Eq. \eqref{formularight} - works under the assumption that $\bm M$ does not have eigenvalues \emph{exactly} equal to $\lambda$. Following the same steps but starting instead from Eq. \eqref{correctfinal}, it would be possible to lift this (mild) restriction by introducing further Fresnel integrals on the r.h.s. We focus here on the simplest case for illustrative purposes. } as
\begin{align}
&\nonumber\mathcal{N}(\lambda)=\lim_{\epsilon\to 0^+}\frac{1}{\pi\mathrm{i}}\left\{\mathrm{Log}\left[\frac{1}{(2\pi)^{N/2}}\int_{(-\infty,\infty)^N}\mathrm{d}\bm x\exp\left[\frac{\mathrm{i}}{2}\bm x^T \left( \lambda\bm{1} -\bm{M}+\mathrm{i}\epsilon\right)\bm x\right]\right]\right.\\
\nonumber &\left. -\mathrm{Log}\left[\frac{1}{(2\pi)^{N/2}}\int_{(-\infty,\infty)^N}\mathrm{d}\bm x\exp\left[-\frac{\mathrm{i}}{2}\bm x^T \left( \lambda\bm{1} -\bm{M}-\mathrm{i}\epsilon\right)\bm x\right]\right]\right\}+\frac{N}{2}\\
&+2\lim_{\epsilon\to 0^+}\left[\Big\lfloor \frac{1}{2}-\frac{1}{2\pi}\mathrm{Im}(\blacksquare)\Big\rfloor-\Big\lfloor \frac{1}{2}-\frac{1}{2\pi}\mathrm{Im}(\blacktriangle)\Big\rfloor\right]\ ,
\end{align}

which can then be rewritten explicitly as
\begin{align}
&\nonumber\mathcal{N}(\lambda)=\lim_{\epsilon\to 0^+}\frac{1}{\pi\mathrm{i}}\left\{\mathrm{Log}\left[\frac{1}{(2\pi)^{N/2}}\int_{(-\infty,\infty)^N}\mathrm{d}\bm x\exp\left[\frac{\mathrm{i}}{2}\bm x^T \left( (\lambda+\mathrm{i}\epsilon)\bm{1} -\bm{M}\right)\bm x\right]\right]\right.\\
&\nonumber\left. -\mathrm{Log}\left[\frac{1}{(2\pi)^{N/2}}\int_{(-\infty,\infty)^N}\mathrm{d}\bm x\exp\left[-\frac{\mathrm{i}}{2}\bm x^T \left( (\lambda-\mathrm{i}\epsilon)\bm{1} -\bm{M}\right)\bm x\right]\right]\right\}+\frac{N}{2}\\
\nonumber &+2\lim_{\epsilon\to 0^+}\left[\Big\lfloor \frac{1}{2}+\frac{\mathcal{N}(\lambda)}{4}-\frac{N}{8} +\frac{1}{4\pi}\sum_k \arctan\left(\frac{\epsilon}{\lambda_k-\lambda}\right)\Big\rfloor-\right.\\
&\left.\Big\lfloor \frac{1}{2}-\frac{\mathcal{N}(\lambda)}{4}+\frac{N}{8} +\frac{1}{4\pi}\sum_k \arctan\left(\frac{-\epsilon}{\lambda_k-\lambda}\right)\Big\rfloor\right].\label{selfcon}
\end{align}

Note that (rather unexpectedly) the $\lambda$-index $\mathcal{N}(\lambda)$ itself has cropped up on the right hand side as well. This is due to as many terms $\pm \pi$ arising in \eqref{crop1} and \eqref{crop2} as there are eigenvalues less than $\lambda$ in the spectrum of $\bm M$. Eq. \eqref{selfcon} is thus to be regarded as a self-consistency equation for $\mathcal{N}(\lambda)$, although at this stage the rather annoying explicit dependence on the eigenvalues has not yet disappeared.

Taking the last limit \emph{inside} the floor brackets is safe, except in the case where $1/2+\mathcal{N}(\lambda)/4-N/8$ (or $1/2-\mathcal{N}(\lambda)/4+N/8$) are integers. This is due to the fact that for $m\in\mathbb{Z}$, the limit and the ``floor" operations do not necessarily commute. Indeed, assume $\lim_{\epsilon\to 0^+} \phi(\epsilon)=0$. Then
\begin{equation}
\lfloor\lim_{\epsilon\to 0^+} (m+\phi(\epsilon))\rfloor = m\ ,
\end{equation}
but
\begin{equation}
\lim_{\epsilon\to 0^+}\lfloor (m+\phi(\epsilon))\rfloor = m\text{ or }m-1\ ,\label{48}
\end{equation}
depending on whether $\phi(\epsilon)>0$ or $<0$ for small positive $\epsilon$.

\emph{Under the assumption that} $1/2+\mathcal{N}(\lambda)/4-N/8$ and $1/2-\mathcal{N}(\lambda)/4+N/8$ are not integers (which needs to be checked case by case), we can safely neglect the arctan terms\footnote{For any finite $N$, the terms $\phi(\epsilon)=\frac{1}{4\pi}\sum_k \arctan\left(\frac{\pm\epsilon}{\lambda_k-\lambda}\right)$ can be made arbitrarily small. In particular, inside a ``floor" bracket $\lfloor x+\phi(\epsilon)\rfloor$, they can be made smaller than the distance between $x$ and its nearest integer (and thus immaterial), unless $x$ is itself an integer.}, and we land on the following self-consistent equation for $\mathcal{N}(\lambda)$
\begin{align}
&\nonumber\mathcal{N}(\lambda)-2\left[\Big\lfloor \frac{1}{2}+\frac{\mathcal{N}(\lambda)}{4}-\frac{N}{8} \Big\rfloor-\Big\lfloor \frac{1}{2}-\frac{\mathcal{N}(\lambda)}{4}+\frac{N}{8} \Big\rfloor\right]-\frac{N}{2}=\\
\nonumber &\lim_{\epsilon\to 0^+}\frac{1}{\pi\mathrm{i}}\left\{\mathrm{Log}\left[\int_{(-\infty,\infty)^N}\mathrm{d}\bm x\exp\left[\frac{\mathrm{i}}{2}\bm x^T \left( (\lambda+\mathrm{i}\epsilon)\bm{1} -\bm{M}\right)\bm x\right]\right]\right.\\
&\left. -\mathrm{Log}\left[\int_{(-\infty,\infty)^N}\mathrm{d}\bm x\exp\left[-\frac{\mathrm{i}}{2}\bm x^T \left(( \lambda-\mathrm{i}\epsilon)\bm{1} -\bm{M}\right)\bm x\right]\right]\right\}\ ,\label{finalformulaselfcons}
\end{align}
where we have also erased the constants $(2\pi)^{-N/2}$ that get cancelled in the difference of Logs. This is one of the main results of this paper.

Note that both integrals on the r.h.s. of \eqref{finalformulaselfcons} are convergent and only depend on the entries of $\bm M$ as desired. Moreover, the r.h.s. is very close to the integral formula that in some papers is claimed to provide directly the $\lambda$-index. This claim, however, can be immediately ruled out for the very same reason why \eqref{formularight} and \eqref{mainformula} are not the same thing: the imaginary part of the difference of two principal logarithms divided by $\pi\mathrm{i}$ is bounded between $-2$ and $2$ and is therefore unfit to ``count" quantities up to $N$.

The way formula \eqref{finalformulaselfcons} becomes operational is as follows: for a given real symmetric matrix $N\times N$ $\bm M$, evaluate the r.h.s. of \eqref{finalformulaselfcons}. Then, for that given value of $N$, create a table of possible values of $\mathcal{N}(\lambda)=0,\ldots,N$, and evaluate the l.h.s. of \eqref{finalformulaselfcons} on each of those. Then, pick the value(s) $\mathcal{N}(\lambda)$ for which equality is reached. For example, if $N=12$, the table of values taken by the l.h.s. for $\mathcal{N}(\lambda)\in\{0,\ldots,12\}$ is
\begin{equation}
\{0, -1, 0, 1, 0, -1, 0, 1, 0, -1, 0, 1, 0\}\ .
\end{equation}
Therefore, if the r.h.s. of \eqref{finalformulaselfcons} comes up $=1$, then the $\lambda$-index of the matrix $\bm M$ could only be $=3$, or $=7$, or $=11$.

\begin{figure}[htb]
\begin{center}
\includegraphics[totalheight=0.4\textheight]{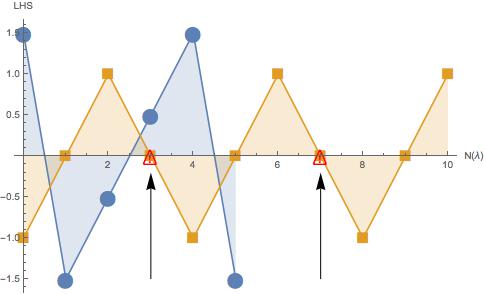}
\caption{Plot of the l.h.s. of \eqref{finalformulaselfcons} as a function of $\mathcal{N}(\lambda)\in [0,\ldots,N]$ for $N=5$ (blue) and $N=10$ (orange). It is evident that several values of the $\lambda$-index may be in principle compatible with a given value of the l.h.s = r.h.s. of \eqref{finalformulaselfcons}. The black arrows point to values of $\mathcal{N}(\lambda)$ for which the assumption that $1/2+\mathcal{N}(\lambda)/4-N/8$ or $1/2-\mathcal{N}(\lambda)/4+N/8$ are not integers is not satisfied. These cases need to be examined separately.}
\label{match}
\end{center}
\end{figure}

As this example already shows, the drawback of the formula is that uniqueness is not guaranteed: in other words, there may exist multiple $\mathcal{N}(\lambda)$ for which equality holds (see also Fig. \ref{match} and Example 2 below). In this sense, the r.h.s. only provides a \emph{set} of \emph{allowed}/\emph{forbidden} values of the index that the matrix $\bm M$ can have, but in general may fail to single out the ``correct" one. This non-uniqueness is the price to pay for wiping the eigenvalues out the game entirely: eq. \eqref{finalformulaselfcons} should be thus regarded as the ``amended'' version\footnote{Bear in mind, however, that outstanding issues still persist around Eq. \eqref{finalformulaselfcons}: it has been derived under the assumptions that (i) the matrix $\bm M$ does not have eigenvalues exactly equal to $\lambda$, and (ii) the arguments of the ``floor" functions are not integers.} of \eqref{mainformula} where $i)$ the ambiguous ``$\log$" is replaced by the principal $\mathrm{Log}$, $ii)$ the inverse square roots of determinants are replaced by multidimensional Fresnel integrals (see also \ref{appD}), but $iii)$ uniqueness of the $\lambda$-index value is generically lost.

For the sake of completeness, we give here below a $4\times 4$ example where uniqueness is eventually attained, and a $5\times 5$ example where it is not.\\
\\
{\bf Example 1.} Take
\begin{equation}
\bm M=\left(
\begin{array}{cccc}
 1 & -2 & 1 & 3 \\
 -2 & -4 & 2 & 2 \\
 1 & 2 & -4 & -5 \\
 3 & 2 & -5 & 3 \\
\end{array}
\right)\ ,
\end{equation}
whose eigenvalues are $\bm\lambda=\{-9.25637,6.54579,-2.46691,1.17749\}$. Therefore $\mathcal{N}(0)=2$. Setting $\bm x =(x_1,x_2,x_3,x_4)^T$, we have
\begin{align}
\nonumber &\bm x^T \left( \lambda\bm{1} -\bm{M}+\mathrm{i}\epsilon\right)\bm x=\bm x^T \left(  -\bm{M}+\mathrm{i}\epsilon\right)\bm x=\\
\nonumber & \mathrm{i} \epsilon (x_1^2+ x_2^2+x_3^2+ x_4^2)-x_1^2+4 x_1 x_2-2 x_1 x_3-6 x_1 x_4+4 x_2^2-4 x_2 x_3\\
&-4 x_2 x_4+4 x_3^2+10 x_3 x_4-3 x_4^2\ ,
\end{align}
and
\begin{align}
\nonumber &\bm x^T \left( \lambda\bm{1} -\bm{M}-\mathrm{i}\epsilon\right)\bm x=\bm x^T \left(  -\bm{M}-\mathrm{i}\epsilon\right)\bm x=\\
\nonumber & -\mathrm{i} \epsilon (x_1^2+ x_2^2+x_3^2+ x_4^2)-x_1^2+4 x_1 x_2-2 x_1 x_3-6 x_1 x_4+4 x_2^2-4 x_2 x_3\\
&-4 x_2 x_4+4 x_3^2+10 x_3 x_4-3 x_4^2\ .
\end{align}

The integrals (evaluated with \textsf{Mathematica} with $\epsilon=10^{-6}$) read
\begin{align}
\int_{(-\infty,\infty)^N}\mathrm{d}\bm x\exp\left[\frac{\mathrm{i}}{2}\bm x^T \left( -\bm{M}+\mathrm{i}\epsilon\right)\bm x\right] &\approx 2.9758\, +7.27041\times 10^{-7}\mathrm{i}\\
\int_{(-\infty,\infty)^N}\mathrm{d}\bm x\exp\left[-\frac{\mathrm{i}}{2}\bm x^T \left( -\bm{M}-\mathrm{i}\epsilon\right)\bm x\right] &\approx 2.9758\, -7.27041\times 10^{-7}\mathrm{i}\ ,
\end{align}
from which the r.h.s. of \eqref{finalformulaselfcons} yields $0$ (to the numerical precision). On the other hand, we can produce the following table for the l.h.s.
of \eqref{finalformulaselfcons} (with $\lambda=0$ and $N=4$)

\begin{table}[h]
\begin{center}
 \begin{tabular}{|c || c c c c c||} 
 \hline
 $\mathcal{N}(\lambda)$ & $0$ & $1$ & $2$ & $3$ & $4$  \\ [0.5ex] 
 \hline\hline
 $1/2 + \mathcal{N}(\lambda)/4 -N/8$ & $0$ & $1/4$ & $1/2$ & $3/4$ & $1$\\
 \hline\hline
  $1/2 - \mathcal{N}(\lambda)/4 +N/8$ & $1$ & $3/4$ & $1/2$ & $1/4$ & $0$\\
 \hline\hline
 l.h.s. of \eqref{finalformulaselfcons} & 0 & -1 & 0 & 1 & 0 \\ [1ex] 
 \hline\hline
\end{tabular}
\end{center}
\caption{Possible values of the l.h.s. of \eqref{finalformulaselfcons} for $\lambda=0$ and $N=4$ depending on the value of $\mathcal{N}(\lambda=0)\in \{0,1,2,3,4\}$.}\label{table1}
\end{table}

One notices that the correct value of the index $\mathcal{N}(0)=2$ produces a l.h.s. that matches the r.h.s. ($=0$). On the other hands, two more incorrect values for the index $\mathcal{N}(0)=0,4$ also appear to be compatible with the r.h.s. . A closer look reveals that in those two cases, the assumption that  $1/2 + \mathcal{N}(\lambda)/4 -N/8$ and/or $1/2 - \mathcal{N}(\lambda)/4 +N/8$ are not integers is not satisfied. Therefore, let us cheat for a moment, and go back to Eq. \eqref{selfcon} to evaluate the limit at $\lambda=0$ and for $\mathcal{N}(0)=0$
\begin{align}
\nonumber &\scalebox{0.98}{$\lim_{\epsilon\to 0^+}\left[\Big\lfloor \frac{1}{2}+\frac{\mathcal{N}(0)}{4}-\frac{N}{8} +\frac{1}{4\pi}\sum_k \arctan\left(\frac{\epsilon}{\lambda_k}\right)\Big\rfloor-\Big\lfloor \frac{1}{2}-\frac{\mathcal{N}(0)}{4}+\frac{N}{8} +\frac{1}{4\pi}\sum_k \arctan\left(\frac{-\epsilon}{\lambda_k}\right)\Big\rfloor\right]=$}\\
&\lim_{\epsilon\to 0^+}\left[\Big\lfloor\frac{1}{4\pi}\sum_k \arctan\left(\frac{\epsilon}{\lambda_k}\right)\Big\rfloor-\Big\lfloor 1 +\frac{1}{4\pi}\sum_k \arctan\left(\frac{-\epsilon}{\lambda_k}\right)\Big\rfloor\right]\label{naive1} 
\end{align}
more carefully. To do so, we clearly need to use some information about the eigenvalues, which - once again - we pledged \emph{not} to use. But please bear with us for a moment.

Having $\mathcal{N}(0)=0$ means that all eigenvalues are positive, therefore $\frac{1}{4\pi}\sum_k \arctan\left(\frac{\epsilon}{\lambda_k}\right)>0$ and $\frac{1}{4\pi}\sum_k \arctan\left(\frac{-\epsilon}{\lambda_k}\right)<0$ for $\epsilon$ positive. Therefore the limit in \eqref{naive1} yields
\begin{equation}
\lim_{\epsilon\to 0^+}[\cdots]=0\ ,
\end{equation}
and not $-1$ as one would have naively obtained by just setting $\epsilon=0$ in \eqref{naive1}. Therefore the l.h.s. of \eqref{finalformulaselfcons} should be corrected and would actually read $-2$ and not $0$, implying that $\mathcal{N}(\lambda=0)=0$ is not a good match anymore.

Similarly, for $\mathcal{N}(\lambda=0)=4$ we need to evaluate the limit 
\begin{align}
\nonumber &\scalebox{0.98}{$\lim_{\epsilon\to 0^+}\left[\Big\lfloor \frac{1}{2}+\frac{\mathcal{N}(0)}{4}-\frac{N}{8} +\frac{1}{4\pi}\sum_k \arctan\left(\frac{\epsilon}{\lambda_k}\right)\Big\rfloor-\Big\lfloor \frac{1}{2}-\frac{\mathcal{N}(0)}{4}+\frac{N}{8} +\frac{1}{4\pi}\sum_k \arctan\left(\frac{-\epsilon}{\lambda_k}\right)\Big\rfloor\right]=$}\\
&\lim_{\epsilon\to 0^+}\left[\Big\lfloor 1+\frac{1}{4\pi}\sum_k \arctan\left(\frac{\epsilon}{\lambda_k}\right)\Big\rfloor-\Big\lfloor \frac{1}{4\pi}\sum_k \arctan\left(\frac{-\epsilon}{\lambda_k}\right)\Big\rfloor\right]\label{naive1a} 
\end{align}
more carefully. Having $\mathcal{N}(\lambda=0)=4$ means that all eigenvalues are negative, therefore $\frac{1}{4\pi}\sum_k \arctan\left(\frac{\epsilon}{\lambda_k}\right)<0$ and $\frac{1}{4\pi}\sum_k \arctan\left(\frac{-\epsilon}{\lambda_k}\right)>0$ for $\epsilon$ positive. Therefore the limits yields
\begin{equation}
\lim_{\epsilon\to 0^+}[\cdots]=0\ ,
\end{equation}
and not $1$ as one would have naively obtained by just setting $\epsilon=0$ in \eqref{naive1a}. Therefore the l.h.s. of \eqref{finalformulaselfcons} should be corrected and would actually read $+2$ and not $0$, implying that $\mathcal{N}(\lambda=0)=4$ is not a good match either.

In summary, the formula \eqref{finalformulaselfcons} (depending only on the entries of the matrix $\bm M$) eventually provides the correct and unique value for the index $\mathcal{N}(\lambda=0)=2$, but this only happens because other potential matchings between the l.h.s. and the r.h.s. can be ruled out following a careful evaluation of the limits in the original formula \eqref{selfcon} (see \eqref{naive1} and \eqref{naive1a}). This strategy - in principle relying on explicit information about eigenvalues, which we should never use - can clearly be followed only in a few rare instances, namely when the sign of the $\arctan$ terms in \eqref{selfcon} can be determined from the mere knowledge of $\mathcal{N}(\lambda)$ alone (e.g. knowing that \emph{all} eigenvalues are positive/negative, without using their explicit values).

Outside these rare instances, the best we can do when either $1/2+\mathcal{N}(\lambda)/4-N/8$ or $1/2-\mathcal{N}(\lambda)/4+N/8$ happen to be integers is to take advantage of Eq. \eqref{48}. In practice, the effect of the $\arctan$ terms - depending explicitly on the exact pattern of eigenvalues - leads \emph{at most} to lowering the ``floor" argument by one. Therefore, the Table \ref{table1} - assuming we could not analyse the ``floor" terms in full detail, as we just did - could be amended as follows.

\begin{table}[h]
\begin{center}
 \begin{tabular}{|c || c c c c c||} 
 \hline
 $\mathcal{N}(\lambda)$ & $0$ & $1$ & $2$ & $3$ & $4$  \\ [0.5ex] 
 \hline\hline
 $1/2 + \mathcal{N}(\lambda)/4 -N/8$ & $0$ & $1/4$ & $1/2$ & $3/4$ & $1$\\
 \hline\hline
First $\lfloor\cdots\rfloor$ in \eqref{selfcon} & $0$ or $-1$ & $1/4$ & $1/2$ & $3/4$ & $1$ or $0$\\
 \hline\hline
  $1/2 - \mathcal{N}(\lambda)/4 +N/8$ & $1$ & $3/4$ & $1/2$ & $1/4$ & $0$\\
 \hline\hline
 Second $\lfloor\cdots\rfloor$ in \eqref{selfcon} & $1$ or $0$ & $1/4$ & $1/2$ & $3/4$ & $0$ or $-1$\\
 \hline\hline
 l.h.s. of \eqref{finalformulaselfcons} & \{0,-2,2\} & -1 & 0 & 1 & \{0,-2,2\} \\ [1ex] 
 \hline\hline
\end{tabular}
\end{center}
\caption{Possible values of the l.h.s. of \eqref{finalformulaselfcons} for $\lambda=0$ and $N=4$ depending on the value of $\mathcal{N}(\lambda=0)\in \{0,1,2,3,4\}$, taking into account possible extra matchings due to the arguments of the ``floor" functions being integers.}\label{table2}
\end{table}
Not being able to analyse the ``floor" functions in full detail clearly has the disappointing consequence that other potential matchings could not be ruled out, and uniqueness would be lost.
{\bf Example 2.} Take now
\begin{equation}
\bm M =\left(
\begin{array}{ccccc}
 -1 & 0.1 & 0 & 0 & 0 \\
 0.1 & -4 & 0 & 0 & 0 \\
 0 & 0 & -5 & 0 & 0 \\
 0 & 0 & 0 & -8 & 0 \\
 0 & 0 & 0 & 0 & -2 
\end{array}
\right)\ ,
\end{equation}
whose eigenvalues are $\bm\lambda=\{-8., -5., -4.00333, -2., -0.99667\}$. Therefore $\mathcal{N}(0)=5$. Setting $\bm x =(x_1,x_2,x_3,x_4,x_5)^T$, we have
\begin{align}
\nonumber &\bm x^T \left( (\lambda+\mathrm{i}\epsilon)\bm{1} -\bm{M}\right)\bm x=\bm x^T \left(  -\bm{M}+\mathrm{i}\epsilon\right)\bm x=\\
& \mathrm{i} \epsilon (x_1^2+ x_2^2+ x_3^2+ x_4^2+x_5^2)+x_1^2-0.2 x_1 x_2+4 x_2^2+5 x_3^2+8 x_4^2+2 x_5^2\ ,
\end{align}
and
\begin{align}
\nonumber &\bm x^T \left( (\lambda-\mathrm{i}\epsilon)\bm{1} -\bm{M}\right)\bm x=\bm x^T \left(  -\bm{M}-\mathrm{i}\epsilon\right)\bm x=\\
& -\mathrm{i} \epsilon (x_1^2+ x_2^2+ x_3^2+ x_4^2+x_5^2)+x_1^2-0.2 x_1 x_2+4 x_2^2+5 x_3^2+8 x_4^2+2 x_5^2\ .
\end{align}

The integrals (evaluated with \textsf{Mathematica} with $\epsilon=10^{-6}$) read
\begin{align}
\int_{(-\infty,\infty)^N}\mathrm{d}\bm x\exp\left[\frac{\mathrm{i}}{2}\bm x^T \left( -\bm{M}+\mathrm{i}\epsilon\right)\bm x\right] &\approx -3.91655-3.91654 \mathrm{i}\\
\int_{(-\infty,\infty)^N}\mathrm{d}\bm x\exp\left[-\frac{\mathrm{i}}{2}\bm x^T \left( -\bm{M}-\mathrm{i}\epsilon\right)\bm x\right] &\approx -3.91655+3.91654 \mathrm{i}\ ,
\end{align}
from which the r.h.s. of \eqref{finalformulaselfcons} yields $-3/2$ (to the numerical precision). On the other hand, we can produce the following table for the l.h.s.
of \eqref{finalformulaselfcons} (with $\lambda=0$ and $N=5$).

\begin{table}[h]
\begin{center}
 \begin{tabular}{|c || c c c c c c||} 
 \hline
 $\mathcal{N}(\lambda)$ & $0$ & $1$ & $2$ & $3$ & $4$ & $5$  \\ [0.5ex] 
 \hline\hline
 $1/2 + \mathcal{N}(\lambda)/4 -N/8$ & $-1/8$ & $1/8$ & $3/8$ & $5/8$ & $7/8$ & $9/8$\\
 \hline\hline
  $1/2 - \mathcal{N}(\lambda)/4 +N/8$ & $9/8$ & $7/8$ & $5/8$ & $3/8$ & $1/8$ & $-1/8$\\
 \hline\hline
 l.h.s. of \eqref{finalformulaselfcons} & $3/2$ & $-3/2$ & $-1/2$ & $1/2$ & $3/2$ & $-3/2$ \\ [1ex] 
 \hline\hline
\end{tabular}
\end{center}
\caption{Possible values of the l.h.s. of \eqref{finalformulaselfcons} for $\lambda=0$ and $N=5$ depending on the value of $\mathcal{N}(\lambda=0)\in \{0,1,2,3,4,5\}$.}\label{table3}
\end{table}

One notices that in this case the correct value for the index $(\mathcal{N}(\lambda=0)=5)$, yielding a l.h.s. equal to the r.h.s. $-3/2$ comes alongside another incorrect matching $\mathcal{N}(\lambda=0)=1$. In this case, the spurious solution of the self-consistency equation \eqref{finalformulaselfcons} cannot be ruled out, because the corresponding arguments of the ``floor" functions are not integers.

\section{Summary}
In summary, we have shown that the claimed \emph{algebraic} identity Eq. \eqref{mainformula} generically fails to count the number of eigenvalues of real symmetric matrices $\bm M$ falling below a threshold $\lambda$ (the $\lambda$-index $\mathcal{N}(\lambda)$) every time the same branch of the complex logarithm is chosen for the two determinants appearing in \eqref{mainformula}. The improved formula \eqref{formularight} is equally unsatisfactory, for two reasons: first, it depends explicitly on the \emph{eigenvalues} of $\bm M$ themselves, and not on its \emph{entries} (making the whole exercise rather pointless); second, it also fails to deal correctly with matrices $\bm M$ having a certain number of eigenvalues \emph{exactly} equal to $\lambda$.

The two determinants in the crippled formula \eqref{mainformula} have been represented as multidimensional Gaussian-Fresnel (or less frequently as Grassmann integrals \cite{cavagna}) in previous studies of the index of random matrices \cite{metz,metzreplica,castillo,fyodorov}. We have shown that these representations \emph{per se} do not cure the problem, which is originated by the phase pattern of complex logarithms.

Indeed, by a very careful evaluation of multidimensional (convergent) Fresnel integrals, supported by several examples and numerical checks, we have demonstrated that 
the difference of principal logarithms of multidimensional Fresnel integrals only yields a set of \emph{possible} $\lambda$-indices of $\bm M$, which may well contain spurious values alongside the correct one. The very same problem would arise if Grassman integrals were used instead of Fresnel.

The inaccurate identity \eqref{mainformula} - and its ``integral" versions - are typically used in connection with the celebrated ``replica trick" \cite{parisi,ej}: it is therefore a very interesting question how the corresponding results turn out to be likely correct while relying on a technically flawed starting point. This issue will be discussed in a separate publication.

\appendix
\section{Complex logarithms}\label{appA}
A complex logarithm is defined as a solution $w$ of the equation
\be
\mathrm{e}^w = z\ ,\label{deflog}
\ee
where $z$ is any nonzero complex number. When $z$ and $w$ are written as $z=r \mathrm{e}^{\mathrm{i}\Theta}$ ($-\pi<\Theta\leq \pi$) and $w=u+\mathrm{i}v$, eq. \eqref{deflog} becomes
\be
\mathrm{e}^u \mathrm{e}^{\mathrm{i}v}=r \mathrm{e}^{\mathrm{i}\Theta}\ .
\ee
It follows that
\be
\mathrm{e}^u=r,\qquad v=\Theta+2n\pi\ ,
\ee
where $n$ is an integer. Therefore
\be
\log z:=w = \ln r +\mathrm{i}(\Theta+2 n\pi),\qquad n=0,\pm 1,\pm2,\ldots\label{deflog2}
\ee
So, $\log z$ is a multi-valued function satisfying $\mathrm{e}^{\log z}=z$ for any $z\neq 0$.

The \emph{principal value} of $\log z$ is the value obtained when $n=0$ and is denoted by $\mathrm{Log}\ z$, that is
\be
\mathrm{Log}\ z=\ln r+\mathrm{i}\Theta\ .
\ee

If $z=r \mathrm{e}^{\mathrm{i}\theta}$ is a nonzero complex number, the argument $\theta$ takes any one of the values $\theta=\Theta+2n\pi$ ($n=0,\pm 1,\pm2,\ldots$) where $\Theta=\mathrm{Arg}\ z$. Therefore, the definition of the logarithmic function \eqref{deflog2} can be written as $\log z=\ln r+\mathrm{i}\theta$.

If we let $\alpha$ denote any real number and restrict the value of $\theta$ so that $\alpha<\theta<\alpha+2\pi$ the function
\be
\log z=\ln r+\mathrm{i}\theta,\qquad r>0,\alpha<\theta<\alpha+2\pi
\ee
is \emph{single-valued} and continuous in the stated domain.

A \emph{branch} of a multiple-valued function $f$ is any single-valued function $F$ that is analytic in some domain at each point $z$ of which the value $F(z)$ is one of the values $f(z)$. The requirement of analyticity, of course, prevents $F$ from taking on a random selection of the values of $f$. Observe that, for each fixed $\alpha$ the single-valued function 
\be
\log z=\ln r+\mathrm{i}\theta,\qquad r>0,\alpha<\theta<\alpha+2\pi
\ee
is a branch of the multiple-valued function 
\be
\log z=\ln r+\mathrm{i}\theta\ .
\ee

Some identities involving \emph{real} logarithms in calculus carry over to complex analysis and others do not (see \cite{haber} for an excellent online resource).

If $z_1$ and $z_2$ denote any two nonzero complex numbers, it is straightforward to show that
\be
\log(z_1 z_2)=\log z_1+\log z_2\label{logsum}
\ee
but this statement involves a multiple-valued function! Hence it means that if two of the three logarithms are specified, then there is a value of the third logarithm such that this equation holds. As noted before, however, the identity \eqref{logsum} is no longer necessarily valid if $\log$ is replaced by $\Log$.

\section{Proof of the improved index formula \eqref{correctfinal}}\label{appC}
The correct formula \eqref{correctfinal} stems from the observation that
\be
\lim_{\epsilon\to 0^+} \left(\Log(\lambda+\mathrm{i}\epsilon-\epsilon)  -  \Log(\lambda+\mathrm{i}\epsilon+\epsilon)\right) = 
\begin{cases}
\mathrm{i}\frac{\pi}{2}&\text{for }\lambda=0\\
0 &\text{otherwise}\ .
\end{cases}
\ee
By definition of principal logarithm
\begin{align}
\Log(\lambda+\mathrm{i}\epsilon-\epsilon)&=\ln\sqrt{(\lambda-\epsilon)^2+\epsilon^2}+\mathrm{i}\ \mathrm{atan2}(\epsilon,\lambda-\epsilon)\\
\Log(\lambda+\mathrm{i}\epsilon+\epsilon)&=\ln\sqrt{(\lambda+\epsilon)^2+\epsilon^2}+\mathrm{i}\ \mathrm{atan2}(\epsilon,\lambda+\epsilon)
\end{align}
where the function $\mathrm{atan2}(y,x)$ is such that $-\pi<\mathrm{atan2}(y,x)<\pi$ and defined in terms of the standard $\arctan(x)$ (whose range is $[-\pi/2,\pi/2]$) as
\be
\mathrm{atan2}(y,x)=\begin{cases}
\arctan(y/x)&\text{for }x>0\\
\arctan(y/x)+\pi&\text{for }y\geq 0,x<0\\
\arctan(y/x)-\pi&\text{for }y<0,x<0\\
+\frac{\pi}{2}&\text{for }y>0,x=0\\
-\frac{\pi}{2}&\text{for }y<0,x=0\\
\text{undefined}&\text{for }y=0,x=0\ .
\end{cases}\label{atan}
\ee
Now there are three cases:
\begin{itemize}
\item $\lambda=0$. In this case $\mathrm{atan2}(\epsilon,-\epsilon)\to\arctan(-1)+\pi=3\pi/4$ for $\epsilon\to 0^+$, while $\mathrm{atan2}(\epsilon,+\epsilon)\to\arctan(1)=\pi/4$. Therefore, 
\be
\lim_{\epsilon\to 0^+} \left(\Log(\lambda+\mathrm{i}\epsilon-\epsilon)  -  \Log(\lambda+\mathrm{i}\epsilon+\epsilon)\right)=\mathrm{i}\frac{\pi}{2}
\ee
as expected.
\item $\lambda>0$. In this case $\mathrm{atan2}(\epsilon,\lambda-\epsilon)\to\arctan\left(\frac{\epsilon}{\lambda-\epsilon}\right)=0$ for $\epsilon\to 0^+$, as well as $\mathrm{atan2}(\epsilon,\lambda+\epsilon)\to\arctan\left(\frac{\epsilon}{\lambda-\epsilon}\right)=0$. Therefore, 
\be
\lim_{\epsilon\to 0^+} \left(\Log(\lambda+\mathrm{i}\epsilon-\epsilon)  -  \Log(\lambda+\mathrm{i}\epsilon+\epsilon)\right)=0
\ee
as expected.
\item $\lambda<0$. In this case $\mathrm{atan2}(\epsilon,\lambda-\epsilon)\to\arctan\left(\frac{\epsilon}{\lambda-\epsilon}\right)+\pi=\pi$ for $\epsilon\to 0^+$, as well as $\mathrm{atan2}(\epsilon,\lambda+\epsilon)\to\arctan\left(\frac{\epsilon}{\lambda-\epsilon}\right)+\pi=\pi$. Therefore, 
\be
\lim_{\epsilon\to 0^+} \left(\Log(\lambda+\mathrm{i}\epsilon-\epsilon)  -  \Log(\lambda+\mathrm{i}\epsilon+\epsilon)\right)=0
\ee
as expected. $\blacksquare$
\end{itemize}
In order to complete the proof of \eqref{correctfinal}, it remains to prove that
\be
\lim_{\epsilon\to 0^+} \left(\Log(\lambda+\mathrm{i}\epsilon)  -  \Log(\lambda-\mathrm{i}\epsilon)\right) = 
\begin{cases}
\mathrm{i}\pi &\text{for }\lambda=0\\
0 &\text{for }\lambda>0\\
2\mathrm{i}\pi &\text{for }\lambda<0\ .
\end{cases}
\ee
We have again by definition of principal logarithm
\begin{align}
\Log(\lambda+\mathrm{i}\epsilon)&=\ln\sqrt{\lambda^2+\epsilon^2}+\mathrm{i}\ \mathrm{atan2}(\epsilon,\lambda)\\
\Log(\lambda-\mathrm{i}\epsilon)&=\ln\sqrt{\lambda^2+\epsilon^2}+\mathrm{i}\ \mathrm{atan2}(-\epsilon,\lambda)\ .
\end{align}

Again there are three cases:
\begin{itemize}
\item $\lambda=0$. In this case $\mathrm{atan2}(\epsilon,0)\to+\pi/2$ for $\epsilon\to 0^+$, while $\mathrm{atan2}(-\epsilon,0)\to -\pi/2$. Therefore, 
\be
\lim_{\epsilon\to 0^+} \left(\Log(\lambda+\mathrm{i}\epsilon)  -  \Log(\lambda-\mathrm{i}\epsilon)\right)=\mathrm{i}\pi
\ee
as expected.
\item $\lambda>0$. In this case $\mathrm{atan2}(\epsilon,\lambda)\to\arctan\left(\frac{\epsilon}{\lambda}\right)=0$ for $\epsilon\to 0^+$, as well as $\mathrm{atan2}(-\epsilon,\lambda)\to\arctan\left(\frac{-\epsilon}{\lambda}\right)=0$. Therefore, 
\be
\lim_{\epsilon\to 0^+} \left(\Log(\lambda+\mathrm{i}\epsilon)  -  \Log(\lambda-\mathrm{i}\epsilon)\right)=0
\ee
as expected.
\item $\lambda<0$. In this case $\mathrm{atan2}(\epsilon,\lambda)\to\arctan\left(\frac{\epsilon}{\lambda}\right)+\pi=\pi$ for $\epsilon\to 0^+$, while $\mathrm{atan2}(-\epsilon,\lambda)\to\arctan\left(\frac{\epsilon}{\lambda}\right)-\pi=-\pi$. Therefore, 
\be
\lim_{\epsilon\to 0^+} \left(\Log(\lambda+\mathrm{i}\epsilon)  -  \Log(\lambda-\mathrm{i}\epsilon)\right)=2\pi\mathrm{i}
\ee
as expected. $\blacksquare$
\end{itemize}

\section{Multidimensional Fresnel integral as an inverse square root of a determinant}\label{appD}

We start from \eqref{identityFresnel}

\begin{equation}
\mathcal{I}_N[\bm M;\lambda;\epsilon]=(2\pi)^{N/2}\exp\left[-\frac{1}{2}\sum_{k=1}^N\mathrm{Log}(\lambda_k-\lambda+\mathrm{i}\epsilon)+\mathrm{i}\frac{N\pi}{4}\right]\ ,\label{identityFresnelapp}
\end{equation}
which can be safely rewritten as
\begin{equation}
\mathcal{I}_N[\bm M;\lambda;\epsilon]=(2\pi)^{N/2} \mathrm{e}^{\mathrm{i}N \pi/4}\prod_{k=1}^N \exp\left[-\frac{1}{2}\mathrm{Log}(\lambda_k-\lambda+\mathrm{i}\epsilon)\right]\ .\label{identityFresnelapp2}
\end{equation}
Next, we use $\exp(a \mathrm{Log}(z))=Z^a$, where $Z^a$ denotes the principal value of the complex number $z^a$ to write
\begin{equation}
\mathcal{I}_N[\bm M;\lambda;\epsilon]=(2\pi)^{N/2} \mathrm{e}^{\mathrm{i}N \pi/4}\prod_{k=1}^N \frac{1}{\sqrt{\lambda_k-\lambda+\mathrm{i}\epsilon}}\ ,\label{identityFresnelapp3}
\end{equation}
where $\sqrt{\cdot}$ denotes the principal square root.

To drag a determinant into the game, we would need to swap the square root and the product symbols. This can only be done at the price of introducing an extra phase, as $(Z_1 Z_2)^a = Z_1^a Z_2^a\mathrm{e}^{2\pi\mathrm{i}a N_+}$, with
\begin{equation}
N_+ =
\begin{cases}
-1 &\text{ if }\mathrm{Arg}\ z_1 +\mathrm{Arg}\ z_2>\pi\\
0 &\text{ if }-\pi<\mathrm{Arg}\ z_1 +\mathrm{Arg}\ z_2\leq\pi\\
1 &\text{ if } \mathrm{Arg}\ z_1 +\mathrm{Arg}\ z_2\leq -\pi\ ,
\end{cases}
\end{equation}
with $\mathrm{Arg}$ denoting the principal argument of $z$. Therefore, we can conclude that
\begin{align}
\nonumber\mathcal{I}_N[\bm M;\lambda;\epsilon] &=(2\pi)^{N/2} \mathrm{e}^{\mathrm{i}N \pi/4}\mathrm{e}^{\mathrm{i}\pi \hat{N}(\bm\lambda)}
\frac{1}{\sqrt{\prod_{k=1}^N (\lambda_k-\lambda+\mathrm{i}\epsilon)}}\\ 
&=(2\pi\mathrm{i})^{N/2}\frac{\mathrm{e}^{\mathrm{i}\pi \hat{N}(\bm\lambda)}}{\sqrt{\det\left[\bm M-\lambda+\mathrm{i}\epsilon\right]}}\ ,\label{identityFresnelapp3}
\end{align}
where we have used $\mathrm{i}^{N/2}=\exp((N/2)\mathrm{Log}(\mathrm{i}))=\exp(\mathrm{i}N\pi/4)$, and $\hat{N}(\bm\lambda)\in\mathbb{Z}$ depends on the specific pattern of phases of the numbers $\{\lambda_k-\lambda+\mathrm{i}\epsilon\}$. The consequence is that a representation of the multidimensional Fresnel integral in terms of inverse (principal) square root of a determinant does exist, but it carries a sign ``ambiguity"\footnote{This is because $\exp(\mathrm{i}\pi m)=\pm 1$, with $m\in\mathbb{Z}$.} that would be resolved knowing the precise phase patterns of the numbers $\{\lambda_k-\lambda+\mathrm{i}\epsilon\}$. However, expressing the integer $ \hat{N}(\bm\lambda)$ in terms of the entries of $\bm M$ alone is rather unnatural, and is preferable to keep adopting the representation \eqref{identityFresnel} that is free of ambiguities.

\vspace{10pt}
{\bf Acknowledgments:} I acknowledge support from the EPSRC Centre for Doctoral Training in Cross-Disciplinary Approaches to Non-Equilibrium Systems (CANES). I am very grateful to Pierfrancesco Urbani, with whom I have worked at the early stage of this project and who has given me very important advice and guidance. I have benefitted from many useful discussion with Yan V Fyodorov, Fabian Aguirre Lopez and Ton Coolen for which I am very grateful.

\end{document}